\documentclass[twoside,a4paper]{article}
\usepackage{xcolor}
\usepackage[backref=section]{hyperref}
\hypersetup{
    colorlinks,
     citecolor=gray,
     filecolor=black,
     linkcolor=darkgray,
     urlcolor=cyan
}

\renewcommand*{\backref}[1]{}
\renewcommand*{\backrefalt}[4]{%
    \ifcase#1%
          \or\mbox{[Cited \S~#2.]}%
          \else\mbox{[Cited \S~#2.]}%
    \fi%
    }
\usepackage{ae} 
\usepackage{aeguill}
\usepackage{amsmath}
\usepackage{amssymb}
\allowdisplaybreaks[4]
\usepackage{longtable}
\newtheorem{mcounter}{mcounter}
\newtheorem{facts}[mcounter]{Facts}

\newtheorem{lemma}[mcounter]{Lemma}
\newtheorem{proposition}[mcounter]{Proposition}

\newtheorem{notation}[mcounter]{Notation}
\newtheorem{notations}[mcounter]{Notations}

\newtheorem{remark}[mcounter]{Remark}
\newtheorem{remarks}[mcounter]{Remarks}

\def\matrixsize#1#2{{({#1}\times{#2})}}

\def\Trace{\textup{Trace}}
\def\Transpose#1{{{}^{\intercal}#1}}
\def\TTrace#1#2#3{\Trace\left({#1}\cdot{}{#2}\cdot{}{\Transpose{#3}}\right)}

\def\TensorRank#1#2#3{{\langle{#1},{#2},{#3}\rangle}}

\def\FMMA#1#2#3#4{{(#1\times{#2}\times{#3}\,;#4)}}
\title{A non-commutative algorithm for multiplying~$\matrixsize{7}{7}$ matrices using~$250$ multiplications}
\author{\href{mailto:Alexandre.Sedoglavic@univ-lille.fr}{Alexandre.Sedoglavic@univ-lille.fr}}
\begin{document}
\maketitle
\begin{abstract}
We present a non-commutative algorithm for multiplying~$\matrixsize{7}{7}$ matrices using~$250$ multiplications and a non-commutative algorithm for multiplying~$\matrixsize{9}{9}$ matrices using~$520$ multiplications. 
These algorithms are obtained using the same divide-and-conquer technique that could be applied to any suitable matrix sizes.
\end{abstract}
\section{Introduction}
The main tool of this note could be summarised in the following proposition:
\begin{proposition}\label{prop:main}--- 
Denoting by~${\langle u,v,w\rangle}$ the number of multiplications necessary to multiply an~$\matrixsize{u}{v}$ matrix with an~$\matrixsize{v}{w}$ matrix to yield an~$\matrixsize{u}{w}$ product matrix, the following relation holds:
  \begin{equation}
    \label{eq:MainResult}
    {\langle u+v,u+v,u+v\rangle}\leq {\langle u,u,u\rangle}+3\,{\langle u,u,v\rangle} + 3\,{\langle v,v,u\rangle} \ \textrm{when}\ u>v.
  \end{equation}
\end{proposition}
For~${(u,v)=(4,3)}$, by selecting already known matrix multiplication algorithms and applying this proposition, we obtain a new upper bounds~$250$ and the explicit corresponding algorithm~\href{http://cristal.univ-lille.fr/~sedoglav/FMM/7x7x7.html}{$\FMMA{7}{7}{7}{250}$}.
\par
In fact, we use the Strassen's matrix multiplication algorithm~\cite{strassen:1969} to \emph{divide} the~$\matrixsize{7}{7}$ matrix multiplication problem into smaller sub-problems; 
the use of three Smirnov's rectangular matrix multiplication algorithms~\cite{smirnov:2013a, Smirnov:2017aa} allows to \emph{conquer} new upper bounds on the number of necessary non-commutative multiplications.
\par
To illustrate this point, we first present a scheme that evaluate the product~${P={N\cdot M}}$:
\begin{equation}
  \label{eq:1}
  N\!=\!\left(\!\!
    \begin{array}{ccc}
      {n_{11}}&{\cdots}&{n_{17}}\\
      {\vdots}&           &{\vdots}\\
      {n_{71}}&{\cdots}&{n_{77}}
    \end{array}\!\!\right)\!,\
  M\!=\!\left(\!\!
    \begin{array}{ccc}
      {m_{11}}&{\cdots}&{m_{17}}\\
      {\vdots}&           &{\vdots}\\
      {m_{71}}&{\cdots}&{m_{77}}
    \end{array}\!\!\right)\!,\
  P\!=\!\left(\!\!
    \begin{array}{ccc}
      {p_{11}}&{\cdots}&{p_{17}}\\
      {\vdots}&           &{\vdots}\\
      {p_{71}}&{\cdots}&{p_{77}}
    \end{array}\!\!\right)\!,
\end{equation}
using~$250$ multiplications.  
This algorithm improves slightly the previous known upper bound~$258$ presented in~\cite{drevet:2011a} and likely obtained with the same kind of techniques presented in this note.
\par
In the last section of this work, we stress the main limitation of our approach by constructing a~$\matrixsize{9}{9}$ matrix multiplication algorithm using~$520$ multiplications (that is only two multiplications less then the corresponding result in~\cite{drevet:2011a} but~$6$ multiplications more then the algorithm~\href{http://cristal.univ-lille.fr/~sedoglav/FMM/9x9x9.html}{$\FMMA{9}{9}{9}{514}$} cited in Table~\ref{table:results}---that summarise what we know about~$286$ matrix multiplication algorithms and was obtained automatically during the elaboration of this note---see~\cite{Sedoglavic:FMMDB} for a more complete list with all details).
This shows that, as the approach presented here is based on the knowledge of fast matrix multiplication algorithms for rectangular matrices, it is limited by the restricted knowledge we have on these algorithms.
\section{Divide}
\label{sec:divide}
For any~$\matrixsize{2}{2}$ matrices:
\begin{equation}
  \label{eq:11}
  {{A={(a_{ij})}_{1\leq i,j\leq 2}},\quad {B={(b_{ij})}_{1\leq i,j\leq
        2}}}\quad \textrm{and}\quad {C={(c_{ij})}_{1\leq i,j\leq 2},}   
\end{equation}
V.~Strassen shows in~\cite{strassen:1969} that the matrix product~${C=A{\cdot}B}$ could be computed by performing the following operations:
\begin{equation}
  \label{eq:StrassenMultiplicationAlgorithm}
  \begin{aligned}
    \begin{aligned}
      t_{1} &= (a_{11} + a_{22}) (b_{11} + b_{22}),   &t_{2} & = (a_{12} - a_{22})(b_{21} + b_{22}), \\
      t_{3} &= (-a_{11} + a_{21}) (b_{11} + b_{12}), &t_{4} & =(a_{11}+a_{12})b_{22},
    \end{aligned}
    \\
    \begin{aligned}
      t_{5} = a_{11} (b_{12} - b_{22}),\ t_{6} = a_{22} (-b_{11} + b_{21}),\ t_{7} = (a_{21} + a_{22}) b_{11},
    \end{aligned}
    \\[\medskipamount]
    \begin{aligned}
      \left(
        \begin{array}{cc}
          c_{11}   &c_{12} \\
          c_{21} &c_{22}
        \end{array}
      \right)= \left(%
        \begin{array}{cc}
          t_{1} + t_{2} - t_{4} + t_{6}  & t_{6} + t_{7}, \\
          t_{4} + t_{5}& t_{1} + t_{3} + t_{5} -t_{7}
        \end{array}\right)\!,
    \end{aligned}
  \end{aligned}
\end{equation}
in the considered non-necessarily commutative coefficients algebra.
\par
To construct our algorithm, we are going to work with the algebra of~$\matrixsize{4}{4}$ matrices and thus, we have to adapt our inputs~${P,N}$ and~$M$~(\ref{eq:1}) to that end. 
So we rewrite these matrices~(\ref{eq:1}) in the following equivalent form:
\begin{equation}
  \label{eq:7}
  X=\left(
    \begin{array}{cccc|ccccc}
      {x_{11}}& x_{12} &x_{13}& 0 & x_{14} & \cdots & {x_{17}}\\
      x_{21} &  x_{22} & x_{23} & 0 & x_{24} & \cdots & x_{27} \\
      {x_{31}}& x_{32} &x_{33}& 0 & x_{34} & \cdots & {x_{37}}\\
      0        &  0 & 0 & 0 & 0 & \cdots & 0 \\\hline
      {x_{41}}& x_{42} &x_{43}& 0 & x_{44} & \cdots & {x_{47}}\\
      \vdots&  \vdots & \vdots & \vdots & \vdots &  & \vdots \\
      {x_{71}}& x_{72} &{x_{73}}& 0 & x_{74} & \cdots & x_{77}
    \end{array}\right)\!,\quad X\in\lbrace{P,M,N}\rbrace,
\end{equation}
in which we have just added a line and a column of zeros.  
After that padding, the product~${P=N\cdot M}$ is unchanged.
\begin{notations}--- 
  Hence, in the sequel~$P$ (resp.~$M,N$) designates the~$\matrixsize{8}{8}$ matrices defined in~(\ref{eq:7}) and~$P_{ij}$ (resp.~$M_{ij}, N_{ij}$) designates~$\matrixsize{4}{4}$ matrices (e.g.~$P_{11}$ stands for the upper left submatrix of~$P$, etc).
\end{notations}
\begin{remark}\label{rem:TensorFramework}--- 
The process of peeling the result (removing rows and columns) of our computations might be better understood if we use---even implicitly---the tensor interpretation of matrix multiplication algorithms. 
Using this framework it appears that the bilinear application~${\mathcal{B}: \mathbb{K}^{\matrixsize{8}{8}}\times \mathbb{K}^{\matrixsize{8}{8}} \mapsto \mathbb{K}^{\matrixsize{8}{8}}}$ with indeterminates~$N$ and~$M$ that defines the matrix multiplication~${\mathcal{B}(N,M)=N\cdot M=P}$ is completely equivalent to the trilinear form~$\mathbb{K}^{\matrixsize{8}{8}}\times \mathbb{K}^{\matrixsize{8}{8}} \times \mathbb{K}^{\matrixsize{8}{8}} \mapsto \mathbb{K}$ with indeterminates~$N,M$ and~$P$ defined by~$\TTrace{N}{M}{P}$ (e.g.\ see~\cite[\S4.6.4, page~506]{Knuth:1997},~\cite[\S~2.5.2]{landsberg:2010},~\cite[\S~2.2]{Benson:2014aa} or~\cite{Dumas:2016aa} for a complete description of this equivalence).
\end{remark}
Computationally, this equivalence induces that the following relation
holds:
\begin{subequations}
  \begin{align}
    \label{eq:8.1}
    \TTrace{N}{M}{P}  & =  \TTrace{\left( {N_{11}}+{N_{22}} \right)}{\left( {M_{11}}+{M_{22}} \right)} {\left( {P_{11}}+{P_{22}} \right)}\\
    \label{eq:8.2}
                      &+ \TTrace{\left( {N_{12}}-{N_{22}} \right)}{\left( {M_{21}}+{M_{22}} \right)}{{P_{11}}}\\
    \label{eq:8.3}
                      &+\TTrace{\left( {N_{21}}-{N_{11}} \right)}{\left( {M_{11}}+{M_{12}} \right)}{{P_{22}}}\\
    \label{eq:8.4}
                      &+ \TTrace{\left( {N_{11}}+{N_{12}} \right) }{  {M_{22}}}{ \left( {P_{21}}-{P_{11}} \right) }\\
    \label{eq:8.5}
                      &+\TTrace{{N_{11}}}{\left( {M_{12}}-{M_{22}} \right) }{ \left( {P_{21}}+{P_{22}}\right)}\\
    \label{eq:8.6}
                      &+\TTrace{{N_{22}}}{\left( {M_{21}}-{M_{11}} \right)}{  \left( {P_{11}}+{P_{12}} \right)}\\
    \label{eq:8.7}
                      &+ \TTrace{\left( {N_{21}}+{N_{22}} \right)}{ {M_{11}}}{\left( {P_{12}}-{P_{22}} \right) }
  \end{align}
\end{subequations}
and as the bilinear application and the trilinear form are equivalent, one could retrieve directly the algorithm from this last form.  
Our original problem is now divided in~$7$ lower dimensional subproblems encoded by trilinear forms. In the next section, we enumerate the algorithms used to compute the matrix products~(\ref{eq:8.1}-\ref{eq:8.7}).
\section{Conquer}
\label{seq:conquer7x7}
The first summand~(\ref{eq:8.1}) involves
unstructured~$\matrixsize{4}{4}$ matrix multiplication that could be
computed using Strassen's algorithm and could be done with~$7^{2}$
multiplications.  Before studying the other summands, we emphasise the
following trivial remarks:
\begin{remarks}\label{rem:1}--- As we consider
  some matrices with zero last column and/or row, recall that the
  product of two~$\matrixsize{4}{4}$
  matrices~${(x_{ij})}_{1\leq i,j \leq 4}$
  and~${(y_{ij})}_{1\leq i,j \leq 4}$ is equivalent to the product of
  the:
  \begin{itemize}
  \item $\matrixsize{4}{3}$
    matrix~${(x_{ij})}_{1\leq i \leq 4, 1\leq j \leq 3}$ with
    the~$\matrixsize{3}{4}$
    matrix~${(y_{ij})}_{1\leq i \leq 3, 1\leq ,j \leq 4}$ when
    for~${1\leq i \leq 4}$ we have~${y_{i4}=0}$ (zero last row);
  \item $\matrixsize{3}{4}$
    matrix~${(x_{ij})}_{1\leq i \leq 3, 1\leq j \leq 4}$ with
    the~$\matrixsize{4}{3}$
    matrix~${(y_{ij})}_{1\leq i \leq 3, 1\leq ,j \leq 4}$ when
    for~${1\leq j \leq 4}$ we have~${y_{4j}=0}$ (zero last column).
  \end{itemize}
\end{remarks}
Let us now review the matrices involved in the
summands~(\ref{eq:8.2}-\ref{eq:8.7}).
\begin{facts}\label{rem:2}--- We notice that by construction:
  \begin{itemize}
  \item the last row and column of~$X_{11}$ are only composed by
    zeros;
  \item $X_{22}$ is a~$\matrixsize{4}{4}$ matrix;
  \item the last column of~${X_{21}-X_{11}}$ is only composed by
    zeros;
  \item the last line of~${X_{11}+X_{12}}$ is only composed by zeros;
  \item ${X_{12}-X_{22}}$ and ${X_{21}-X_{22}}$
    are~$\matrixsize{4}{4}$ matrices without zero row or column.
  \end{itemize}
\end{facts}
Hence, taking into account Remarks~\ref{rem:1} and Facts~\ref{rem:2},
we have a better description of the sub-problems considered in this
section:
\begin{remarks}\label{rem:subproblems}--- The summand:
  \begin{itemize}
  \item (\ref{eq:8.2}) involves an~$\matrixsize{3}{4}$
    times~$\matrixsize{4}{3}$ times~$\matrixsize{3}{3}$ matrices
    product;
  \item (\ref{eq:8.3}) involves an~$\matrixsize{4}{3}$
    times~$\matrixsize{3}{4}$ times~$\matrixsize{4}{4}$ matrices
    product;
  \item (\ref{eq:8.4}) involves an~$\matrixsize{3}{4}$
    times~$\matrixsize{4}{4}$ times~$\matrixsize{4}{3}$ matrices
    product;
  \item (\ref{eq:8.5}) involves an~$\matrixsize{3}{3}$
    times~$\matrixsize{3}{4}$ times~$\matrixsize{4}{3}$ matrices
    product;
  \item (\ref{eq:8.6}) involves an~$\matrixsize{4}{4}$
    times~$\matrixsize{4}{3}$ times~$\matrixsize{3}{4}$ matrices
    product;
  \item (\ref{eq:8.7}) involves an~$\matrixsize{4}{3}$
    times~$\matrixsize{3}{3}$ times~$\matrixsize{3}{4}$ matrices
    product.
  \end{itemize}
\end{remarks}
\begin{remark}\label{rem:knuth}--- We also rely our construction on
  the representation of matrix multiplication algorithm by trilinear
  forms and the underlying tensor representation (see
  Remark~\ref{rem:TensorFramework}) because, as quoted
  in~\cite[\S~4.6.4~p.~507]{Knuth:1997}:
  \begin{quote}
    ``[$\ldots$], a normal scheme for evaluating
    an~$\matrixsize{m}{n}$ times~$\matrixsize{n}{s}$ matrix product
    implies the existence of a normal scheme to evaluate
    an~$\matrixsize{n}{s}$ times~$\matrixsize{s}{m}$ matrix product
    using the same number of chain multiplications.''
  \end{quote}
\end{remark}
This is exactly what we are using in the sequel. 
To do so and in order to express complexity of the summands~(\ref{eq:8.2}-\ref{eq:8.7}), we (re)introduce more precisely the following notations (already used in Proposition~\ref{prop:main}):
\begin{notation}\label{not:rank}--- For matrices~$U,V$ and~$W$ of
  size~$\matrixsize{u}{v}, \matrixsize{v}{w}$ and~$\matrixsize{w}{u}$,
  we denote by~${\langle u,v,w\rangle}$ the known supremum on the
  multiplication necessary for
  computing~${\TTrace{U}{V}{W}}$ (that is the number
  of multiplication used by the best known algorithm allowing to
  compute~${{U\cdot V}=W}$ (a.k.a.\ tensor rank)).
\end{notation}
\begin{remarks}--- Using this notation, we see that Remarks~\ref{rem:subproblems} can be restated as follow:
  \begin{itemize}
  \item (\ref{eq:8.2}) can be computed using~${\langle 3,4,3\rangle}$
    multiplications;
  \item (\ref{eq:8.3}) can be computed using~${\langle 4,3,4\rangle}$
    multiplications;
  \item (\ref{eq:8.4}) can be computed using~${\langle 3,4,4\rangle}$
    multiplications;
  \item (\ref{eq:8.5}) can be computed using~${\langle 3,3,4\rangle}$
    multiplications;
  \item (\ref{eq:8.6}) can be computed using~${\langle 4,4,3\rangle}$
    multiplications;
  \item (\ref{eq:8.7}) can be computed using~${\langle 4,3,3\rangle}$
    multiplications.
  \end{itemize}
\end{remarks}
We are going to see that we need only two algorithms to perform all these computations.  
In fact, the Remark~\ref{rem:knuth} is a direct consequence of the following~$\Trace$ properties:
\begin{equation}
  \label{eq:12.1}
  \begin{aligned}
    {\Trace\left(U\cdot V\cdot W\right)} &={\Trace\left(W\cdot U\cdot
        V\right)}&&= {\Trace\left(V\cdot W\cdot U\right)}, \\
    &= \Trace\big( \Transpose{\!\left(U\cdot V\cdot W\right)}\big) &&=
    \Trace\big( \Transpose{W} \cdot\Transpose{V}\cdot\Transpose{U}
    \big).
  \end{aligned}
\end{equation}
To be more precise, Relations~(\ref{eq:12.1}) imply the following
well-known result:
\begin{lemma}--- The following relations hold:
  \begin{equation}
    \label{eq:9}
    {\langle u,v,w\rangle}={\langle w,u,v\rangle}={\langle v,w,u\rangle}
    ={\langle v,u,w\rangle}={\langle w,v,u\rangle}={\langle u,w,v\rangle}.
  \end{equation}
\end{lemma}
This allows us to state that algorithm~(\ref{eq:8.2}-\ref{eq:8.7})
requires:
\begin{equation}
  \label{eq:10}
  {\langle 4,4,4\rangle} + 3\,{\langle 3,3,4\rangle}+3\,{\langle 3,4,4\rangle}
\end{equation}
multiplications. As A.~V.~Smirnov states
that~${\langle 3,3,4\rangle=29}$ and~${\langle 3,4,4\rangle=38}$
in~\cite[Table~1,~N$\textsuperscript{\tiny\underline{o}}$~13
and~21]{smirnov:2013a}, we conclude that our algorithm
required~${250}$ multiplications (${7^{2}+3\cdot 29+3\cdot
  38}$). Furthermore, Smirnov provides in~\cite{Smirnov:2017aa} the
explicit description of these algorithms, allowing us to do the same
with the first algorithm constructed in this note at url:
\begin{center}
\url{http://cristal.univ-lille.fr/~sedoglav/FMM/7x7x7.html}.
\end{center}
In the next section, we show how to apply the very same manipulations
to the product of two~$\matrixsize{9}{9}$ matrices.
\section{An algorithm for multiplying~$\matrixsize{9}{9}$ matrices}
As in the previous section, we are going to pad
our~$\matrixsize{9}{9}$ matrices:
\begin{equation}
  \label{eq:Input9x9Matrices}
  {{N={(n_{ij})}_{1\leq i,j\leq 9}},\quad {M={(M_{ij})}_{1\leq i,j\leq
        9}}}\quad \textrm{and}\quad {P={(P_{ij})}_{1\leq i,j\leq 9},}   
\end{equation}
in order to work this time with equivalent~$\matrixsize{12}{12}$
matrices:
\begin{equation}
  \label{eq:Padded9x9}
  Y=\left(
    \begin{array}{cccccc|cccccc}
      {y_{11}}& y_{12} &y_{13}& 0 & 0 & 0 & y_{14} & \cdots & {y_{19}}\\
      y_{21} &  y_{22} & y_{23} & 0 & 0 & 0& y_{24} & \cdots & y_{29} \\
      {y_{31}}& y_{32} &y_{33}& 0 & 0 & 0& y_{34} & \cdots & {y_{39}}\\
      0        &  0 & 0 & 0 & 0 & 0 & 0 & \cdots & 0 \\
      0        &  0 & 0 & 0 & 0 & 0 & 0 & \cdots & 0 \\
      0        &  0 & 0 & 0 & 0 & 0 & 0 & \cdots & 0 \\\hline
      {y_{41}}& y_{42} &y_{43}& 0 &0 & 0 & y_{44} & \cdots & {y_{49}}\\
      \vdots&  \vdots & \vdots & \vdots & \vdots & \vdots &\vdots &  & \vdots \\
      {y_{91}}& y_{92} &{y_{93}}& 0 & 0 & 0 & y_{94} & \cdots & y_{99}
    \end{array}\right)\!,\quad Y\in\lbrace{P,M,N}\rbrace,
\end{equation}
After that padding, the product~${P=N\cdot M}$ is unchanged.
\begin{notations}--- In the sequel~$P$ (resp.~$M,N$) designates
  the~$\matrixsize{12}{12}$ matrices defined in~(\ref{eq:Padded9x9})
  and~$P_{ij}$ (resp.~$M_{ij}, N_{ij}$) designates~$\matrixsize{6}{6}$
  matrices (e.g.~$P_{11}$ stands for the upper left submatrix of~$P$,
  etc).
\end{notations}
The process described in Section~\ref{seq:conquer7x7}
remains---mutatis mutandis---exactly the same and we obtain the
following special case of Proposition~\ref{prop:main}:
\begin{lemma}--- With the Notations~\ref{not:rank}, there exists an
  algorithm that multiply two~$\matrixsize{9}{9}$ matrices
  using~${ {\langle 6,6,6\rangle} + 3\,{\langle
      6,6,3\rangle}+3\,{\langle 6,3,3\rangle}}$ multiplications.
\end{lemma}
Again, we found
in~\cite[\mbox{Table~1,~N$\textsuperscript{\tiny\underline{o}}$~27}]{smirnov:2013a}
that~${\langle 3,3,6\rangle=40}$ (an explicit form of this algorithm
could be found in the source code of~\cite{Benson:2014aa} or in the collection~\cite{Sedoglavic:FMMDB}).  But this
time, we do not have found in the literature any \emph{fast} matrix
multiplication algorithm for~${\langle 6,6,3\rangle}$ and we have to
provide an algorithm for~${\langle 6,6,6\rangle}$.  Nevertheless, as
done implicitly in Section~\ref{sec:divide}, we are going to use again
the following constructive result on tensor's Kronecker product.
\begin{lemma}\label{lem:TensorKroneckerProduct}--- Given an algorithm
  computing~$\matrixsize{u_{1}}{v_{1}}$
  times~$\matrixsize{v_{1}}{w_{1}}$ matrix product
  using~${\langle u_{1},v_{1},w_{1}\rangle}$ multiplications and an
  algorithm computing~$\matrixsize{u_{2}}{v_{2}}$
  times~$\matrixsize{v_{2}}{w_{2}}$ matrix product
  using~${\langle u_{2},v_{2},w_{2}\rangle}$ multiplications, one can
  construct an algorithm
  computing~$\matrixsize{u_{1}u_{2}}{v_{1}v_{2}}$
  by~$\matrixsize{v_{1}v_{2}}{w_{1}w_{2}}$ matrix multiplication
  using~${{\langle u_{1},v_{1},w_{1}\rangle}\cdot {\langle
      u_{2},v_{2},w_{2}\rangle}}$ multiplications (a.k.a.\ the
  tensor's Kronecker product of the two previous algorithms).
\end{lemma}
Hence, as we know that trivially~${{\langle 1,2,2\rangle}=4}$, we
conclude that~${{\langle 6,6,6\rangle}}$ is equal
to~${{\langle 6,3,3\rangle}\cdot {\langle 1,2,2\rangle}}$ (that
is~$160$) and that~${{\langle 6,6,3\rangle}=80}$.  So, the algorithm
constructed in this section requires~${520}$ multiplications
(${4\cdot 40+3\cdot(40\cdot 2)+3\cdot 40}$). 
\section{Concluding remarks}
The complexity of our algorithm for multiplying~$\matrixsize{9}{9}$
matrices could likely be improved by finding a \emph{better} algorithm
for~${\langle 6,6,3\rangle}$ and~${\langle 6,6,6\rangle}$ then those
used above (algorithms obtained by tensor Kronecker product are not
always optimal as shown by the fact
that~${{\langle 3,3,3\rangle}\cdot{\langle 3,3,3\rangle}}$ is equal
to~$529$ while computations summarised in Table~\ref{table:results}
show that~\href{http://cristal.univ-lille.fr/~sedoglav/FMM/9x9x9.html}{${\langle 9,9,9\rangle}$} is now~$514$).
\par
By combining tensor's based description of matrix multiplication
algorithms with rectangular algorithms found by numerical computer
search (see~\mbox{\cite[\S~2.3.2]{Benson:2014aa}}
and~\cite{smirnov:2013a}), it is possible---as already shown
in~\cite{drevet:2011a}---to improve the theoretical complexity of
small size matrix products (see Table~\ref{table:results} in appendix
in which new results obtained with the method presented in this note
are in bold face).
\par
The author thinks that some symmetry-based geometrical methods could
reduce further the upper bounds presented in this note.
\bibliographystyle{acm}

\appendix
\section{Summary of known results}
We gather below a summary of some known results up to~$\TensorRank{12}{12}{12}$.
\label{table:results}
\begin{longtable}[c]{cr|llr}
  $\TensorRank{1}{x}{y}$ & ${x\cdot y}$ & ${x\cdot y}$ &  trivial algorithms  \\
  $\TensorRank{2}{2}{2}$&$7$&$8$&Strassen~\cite{strassen:1969}\\
$\TensorRank{2}{3}{3}$&$15$&$18$&Hopcroft \& Kerr~\cite{hopcroft:1971}\\
$\TensorRank{2}{3}{4}$&$20$&$24$&Hopcroft \& Kerr~\cite{hopcroft:1971}\\
$\TensorRank{2}{3}{5}$&$25$&$30$&Hopcroft \& Kerr~\cite{hopcroft:1971}\\
$\TensorRank{2}{3}{6}$&$\mathrm{30}$&$36$&${\TensorRank{2}{3}{3}\cdot\TensorRank{1}{1}{2}}$\\
$\TensorRank{2}{3}{7}$&$\mathrm{35}$&$42$&${\TensorRank{2}{3}{3}+\TensorRank{2}{3}{4}}$\\
$\TensorRank{2}{3}{8}$&$\mathrm{40}$&$48$&${\TensorRank{2}{3}{4}\cdot\TensorRank{1}{1}{2}}$\\
$\TensorRank{2}{3}{9}$&$\mathrm{45}$&$54$&${\TensorRank{2}{3}{3}\cdot\TensorRank{1}{1}{3}}$\\
$\TensorRank{2}{3}{10}$&$\mathrm{50}$&$60$&${\TensorRank{2}{3}{4}+\TensorRank{2}{3}{6}}$\\
$\TensorRank{2}{3}{11}$&$\mathrm{55}$&$66$&${\TensorRank{2}{3}{7}+\TensorRank{2}{3}{4}}$\\
$\TensorRank{2}{3}{12}$&$\mathrm{60}$&$72$&${\TensorRank{1}{1}{2}\cdot\TensorRank{2}{3}{6}}$\\
$\TensorRank{2}{4}{4}$&$26$&$32$&Hopcroft \& Kerr~\cite{hopcroft:1971}\\
$\TensorRank{2}{4}{5}$&$33$&$40$&Hopcroft \& Kerr~\cite{hopcroft:1971}\\
$\TensorRank{2}{4}{6}$&$39$&$48$&Hopcroft \& Kerr~\cite{hopcroft:1971}\\
$\TensorRank{2}{4}{7}$&$\mathrm{46}$&$56$&${\TensorRank{2}{4}{3}+\TensorRank{2}{4}{4}}$\\
$\TensorRank{2}{4}{8}$&$\mathrm{52}$&$64$&${\TensorRank{2}{4}{4}\cdot\TensorRank{1}{1}{2}}$\\
$\TensorRank{2}{4}{9}$&$\mathrm{59}$&$72$&${\TensorRank{2}{4}{4}+\TensorRank{2}{4}{5}}$\\
$\TensorRank{2}{4}{10}$&$\mathrm{65}$&$80$&${\TensorRank{2}{4}{4}+\TensorRank{2}{4}{6}}$\\
$\TensorRank{2}{4}{11}$&$\mathrm{72}$&$88$&${\TensorRank{2}{4}{7}+\TensorRank{2}{4}{4}}$\\
$\TensorRank{2}{4}{12}$&$\mathrm{78}$&$96$&${\TensorRank{1}{1}{3}\cdot\TensorRank{2}{4}{4}}$\\
$\TensorRank{2}{5}{5}$&$40$&$50$&Hopcroft \& Kerr~\cite{hopcroft:1971}\\
$\TensorRank{2}{5}{6}$&$48$&$60$&Hopcroft \& Kerr~\cite{hopcroft:1971}\\
$\TensorRank{2}{5}{7}$&$56$&$70$&Hopcroft \& Kerr~\cite{hopcroft:1971}\\
$\TensorRank{2}{5}{8}$&$64$&$80$&Hopcroft \& Kerr~\cite{hopcroft:1971}\\
$\TensorRank{2}{5}{9}$&$72$&$90$&Hopcroft \& Kerr~\cite{hopcroft:1971}\\
$\TensorRank{2}{5}{10}$&$\mathrm{80}$&$100$&${\TensorRank{2}{5}{5}\cdot\TensorRank{1}{1}{2}}$\\
$\TensorRank{2}{5}{11}$&$\mathrm{88}$&$110$&${\TensorRank{2}{5}{6}+\TensorRank{2}{5}{5}}$\\
$\TensorRank{2}{5}{12}$&$\mathrm{96}$&$120$&${\TensorRank{2}{5}{6}+\TensorRank{2}{5}{6}}$\\
$\TensorRank{2}{6}{6}$&$57$&$72$&Hopcroft \& Kerr~\cite{hopcroft:1971}\\
$\TensorRank{2}{6}{7}$&$67$&$84$&Hopcroft \& Kerr~\cite{hopcroft:1971}\\
$\TensorRank{2}{6}{8}$&$76$&$96$&Hopcroft \& Kerr~\cite{hopcroft:1971}\\
$\TensorRank{2}{6}{9}$&$\mathrm{87}$&$108$&${\TensorRank{2}{6}{4}+\TensorRank{2}{6}{5}}$\\
$\TensorRank{2}{6}{10}$&$95$&$120$&Hopcroft \& Kerr~\cite{hopcroft:1971}\\
$\TensorRank{2}{6}{11}$&$\mathrm{105}$&$132$&${\TensorRank{2}{6}{5}+\TensorRank{2}{6}{6}}$\\
$\TensorRank{2}{6}{12}$&$\mathrm{114}$&$144$&${\TensorRank{1}{1}{2}\cdot\TensorRank{2}{6}{6}}$\\
$\TensorRank{2}{7}{7}$&$77$&$98$&Hopcroft \& Kerr~\cite{hopcroft:1971}\\
$\TensorRank{2}{7}{8}$&$\mathrm{91}$&$112$&${\TensorRank{2}{7}{3}+\TensorRank{2}{7}{5}}$\\
$\TensorRank{2}{7}{9}$&$99$&$126$&Hopcroft \& Kerr~\cite{hopcroft:1971}\\
$\TensorRank{2}{7}{10}$&$110$&$140$&Hopcroft \& Kerr~\cite{hopcroft:1971}\\
$\TensorRank{2}{7}{11}$&$121$&$154$&Hopcroft \& Kerr~\cite{hopcroft:1971}\\
$\TensorRank{2}{7}{12}$&$132$&$168$&Hopcroft \& Kerr~\cite{hopcroft:1971}\\
$\TensorRank{2}{8}{8}$&$100$&$128$&Hopcroft \& Kerr~\cite{hopcroft:1971}\\
$\TensorRank{2}{8}{9}$&$\mathrm{116}$&$144$&${\TensorRank{2}{8}{4}+\TensorRank{2}{8}{5}}$\\
$\TensorRank{2}{8}{10}$&$\mathrm{128}$&$160$&${\TensorRank{1}{1}{2}\cdot\TensorRank{2}{8}{5}}$\\
$\TensorRank{2}{8}{11}$&$\mathrm{140}$&$176$&${\TensorRank{2}{8}{3}+\TensorRank{2}{8}{8}}$\\
$\TensorRank{2}{8}{12}$&$\mathrm{152}$&$192$&${\TensorRank{2}{8}{4}+\TensorRank{2}{8}{8}}$\\
$\TensorRank{2}{9}{9}$&$126$&$162$&Hopcroft \& Kerr~\cite{hopcroft:1971}\\
$\TensorRank{2}{9}{10}$&$\mathrm{144}$&$180$&${\TensorRank{1}{1}{2}\cdot\TensorRank{2}{9}{5}}$\\
$\TensorRank{2}{9}{11}$&$\mathrm{158}$&$198$&${\TensorRank{2}{9}{2}+\TensorRank{2}{9}{9}}$\\
$\TensorRank{2}{9}{12}$&$168$&$216$&Hopcroft \& Kerr~\cite{hopcroft:1971}\\
$\TensorRank{2}{10}{10}$&$155$&$200$&Hopcroft \& Kerr~\cite{hopcroft:1971}\\
$\TensorRank{2}{10}{11}$&$\mathrm{175}$&$220$&${\TensorRank{2}{10}{1}+\TensorRank{2}{10}{10}}$\\
$\TensorRank{2}{10}{12}$&$\mathrm{190}$&$240$&${\TensorRank{2}{10}{2}+\TensorRank{2}{10}{10}}$\\
$\TensorRank{2}{11}{11}$&$187$&$242$&Hopcroft \& Kerr~\cite{hopcroft:1971}\\
$\TensorRank{2}{11}{12}$&$\mathrm{209}$&$264$&${\TensorRank{2}{11}{1}+\TensorRank{2}{11}{11}}$\\
$\TensorRank{2}{12}{12}$&$222$&$288$&Hopcroft \& Kerr~\cite{hopcroft:1971}\\
$\TensorRank{3}{3}{3}$&$23$&$27$&Laderman~\cite{laderman:1976a}\\
$\TensorRank{3}{3}{4}$&$29$&$36$&Smirnov~\cite{smirnov:2013a}\\
$\TensorRank{3}{3}{5}$&$36$&$45$&Smirnov~\cite{smirnov:2013a}\\
$\TensorRank{3}{3}{6}$&$40$&$54$&Smirnov~\cite{smirnov:2013a}\\
$\TensorRank{3}{3}{7}$&$\mathrm{49}$&$63$&${\TensorRank{3}{3}{1}+\TensorRank{3}{3}{6}}$\\
$\TensorRank{3}{3}{8}$&$\mathrm{55}$&$72$&${\TensorRank{3}{3}{6}+\TensorRank{3}{3}{2}}$\\
$\TensorRank{3}{3}{9}$&$\mathrm{63}$&$81$&${\TensorRank{3}{3}{3}+\TensorRank{3}{3}{6}}$\\
$\TensorRank{3}{3}{10}$&$\mathrm{69}$&$90$&${\TensorRank{3}{3}{4}+\TensorRank{3}{3}{6}}$\\
$\TensorRank{3}{3}{11}$&$\mathrm{76}$&$99$&${\TensorRank{3}{3}{5}+\TensorRank{3}{3}{6}}$\\
$\TensorRank{3}{3}{12}$&$\mathrm{80}$&$108$&${\TensorRank{1}{1}{2}\cdot\TensorRank{3}{3}{6}}$\\
$\TensorRank{3}{4}{4}$&$38$&$48$&Smirnov~\cite{smirnov:2013a}\\
$\TensorRank{3}{4}{5}$&$48$&$60$&Smirnov~\cite{smirnov:2013a}\\
$\TensorRank{3}{4}{6}$&$\mathrm{58}$&$72$&${\TensorRank{3}{4}{3}\cdot\TensorRank{1}{1}{2}}$\\
$\TensorRank{3}{4}{7}$&$\mathrm{67}$&$84$&${\TensorRank{3}{4}{3}+\TensorRank{3}{4}{4}}$\\
$\TensorRank{3}{4}{8}$&$\mathrm{76}$&$96$&${\TensorRank{3}{4}{4}\cdot\TensorRank{1}{1}{2}}$\\
$\TensorRank{3}{4}{9}$&$\mathrm{86}$&$108$&${\TensorRank{3}{4}{4}+\TensorRank{3}{4}{5}}$\\
$\TensorRank{3}{4}{10}$&$\mathrm{96}$&$120$&${\TensorRank{3}{4}{2}+\TensorRank{3}{4}{8}}$\\
$\TensorRank{3}{4}{11}$&$\mathrm{105}$&$132$&${\TensorRank{3}{4}{3}+\TensorRank{3}{4}{8}}$\\
$\TensorRank{3}{4}{12}$&$\mathrm{114}$&$144$&${\TensorRank{1}{1}{3}\cdot\TensorRank{3}{4}{4}}$\\
$\TensorRank{3}{5}{5}$&$\mathrm{61}$&$75$&${\TensorRank{3}{5}{2}+\TensorRank{3}{5}{3}}$\\
$\TensorRank{3}{5}{6}$&$\mathrm{70}$&$90$&${\TensorRank{3}{2}{6}+\TensorRank{3}{3}{6}}$\\
$\TensorRank{3}{5}{7}$&$\mathrm{84}$&$105$&${\TensorRank{3}{2}{7}+\TensorRank{3}{3}{7}}$\\
$\TensorRank{3}{5}{8}$&$\mathrm{95}$&$120$&${\TensorRank{3}{5}{6}+\TensorRank{3}{5}{2}}$\\
$\TensorRank{3}{5}{9}$&$\mathrm{106}$&$135$&${\TensorRank{3}{5}{3}+\TensorRank{3}{5}{6}}$\\
$\TensorRank{3}{5}{10}$&$\mathrm{118}$&$150$&${\TensorRank{3}{5}{4}+\TensorRank{3}{5}{6}}$\\
$\TensorRank{3}{5}{11}$&$\mathbf{130}$&$165$&${2\,\TensorRank{2}{3}{5}+2\,\TensorRank{3}{3}{6}}$\\
$\TensorRank{3}{5}{12}$&$\mathbf{140}$&$180$&${2\,\TensorRank{2}{3}{6}+2\,\TensorRank{3}{3}{6}}$\\
$\TensorRank{3}{6}{6}$&$\mathrm{80}$&$108$&${\TensorRank{3}{3}{6}\cdot\TensorRank{1}{2}{1}}$\\
$\TensorRank{3}{6}{7}$&$\mathrm{98}$&$126$&${\TensorRank{3}{3}{7}\cdot\TensorRank{1}{2}{1}}$\\
$\TensorRank{3}{6}{8}$&$\mathrm{110}$&$144$&${\TensorRank{1}{2}{1}\cdot\TensorRank{3}{3}{8}}$\\
$\TensorRank{3}{6}{9}$&$\mathrm{120}$&$162$&${\TensorRank{3}{6}{3}\cdot\TensorRank{1}{1}{3}}$\\
$\TensorRank{3}{6}{10}$&$\mathrm{138}$&$180$&${\TensorRank{3}{6}{1}+\TensorRank{3}{6}{9}}$\\
$\TensorRank{3}{6}{11}$&$\mathrm{150}$&$198$&${\TensorRank{3}{6}{2}+\TensorRank{3}{6}{9}}$\\
$\TensorRank{3}{6}{12}$&$\mathrm{160}$&$216$&${\TensorRank{1}{1}{2}\cdot\TensorRank{3}{6}{6}}$\\
$\TensorRank{3}{7}{7}$&$\mathrm{116}$&$147$&${\TensorRank{3}{7}{3}+\TensorRank{3}{7}{4}}$\\
$\TensorRank{3}{7}{8}$&$\mathrm{131}$&$168$&${\TensorRank{3}{3}{8}+\TensorRank{3}{4}{8}}$\\
$\TensorRank{3}{7}{9}$&$\mathrm{147}$&$189$&${\TensorRank{3}{7}{3}\cdot\TensorRank{1}{1}{3}}$\\
$\TensorRank{3}{7}{10}$&$\mathrm{165}$&$210$&${\TensorRank{3}{7}{4}+\TensorRank{3}{7}{6}}$\\
$\TensorRank{3}{7}{11}$&$\mathrm{180}$&$231$&${\TensorRank{3}{7}{3}+\TensorRank{3}{7}{8}}$\\
$\TensorRank{3}{7}{12}$&$\mathrm{194}$&$252$&${\TensorRank{3}{3}{12}+\TensorRank{3}{4}{12}}$\\
$\TensorRank{3}{8}{8}$&$\mathrm{150}$&$192$&${\TensorRank{3}{8}{2}+\TensorRank{3}{8}{6}}$\\
$\TensorRank{3}{8}{9}$&$\mathrm{165}$&$216$&${\TensorRank{1}{1}{3}\cdot\TensorRank{3}{8}{3}}$\\
$\TensorRank{3}{8}{10}$&$\mathrm{186}$&$240$&${\TensorRank{3}{8}{4}+\TensorRank{3}{8}{6}}$\\
$\TensorRank{3}{8}{11}$&$\mathrm{205}$&$264$&${\TensorRank{3}{8}{2}+\TensorRank{3}{8}{9}}$\\
$\TensorRank{3}{8}{12}$&$\mathrm{220}$&$288$&${\TensorRank{1}{1}{4}\cdot\TensorRank{3}{8}{3}}$\\
$\TensorRank{3}{9}{9}$&$\mathrm{183}$&$243$&${\TensorRank{3}{9}{3}+\TensorRank{3}{9}{6}}$\\
$\TensorRank{3}{9}{10}$&$\mathrm{206}$&$270$&${\TensorRank{3}{9}{4}+\TensorRank{3}{9}{6}}$\\
$\TensorRank{3}{9}{11}$&$\mathrm{226}$&$297$&${\TensorRank{3}{9}{5}+\TensorRank{3}{9}{6}}$\\
$\TensorRank{3}{9}{12}$&$\mathrm{240}$&$324$&${\TensorRank{1}{1}{2}\cdot\TensorRank{3}{9}{6}}$\\
$\TensorRank{3}{10}{10}$&$\mathrm{234}$&$300$&${\TensorRank{3}{10}{4}+\TensorRank{3}{10}{6}}$\\
$\TensorRank{3}{10}{11}$&$\mathrm{255}$&$330$&${\TensorRank{3}{10}{3}+\TensorRank{3}{10}{8}}$\\
$\TensorRank{3}{10}{12}$&$\mathrm{274}$&$360$&${\TensorRank{3}{4}{12}+\TensorRank{3}{6}{12}}$\\
$\TensorRank{3}{11}{11}$&$\mathrm{280}$&$363$&${\TensorRank{3}{11}{5}+\TensorRank{3}{11}{6}}$\\
$\TensorRank{3}{11}{12}$&$\mathrm{300}$&$396$&${\TensorRank{3}{11}{6}+\TensorRank{3}{11}{6}}$\\
$\TensorRank{3}{12}{12}$&$\mathrm{320}$&$432$&${\TensorRank{1}{2}{2}\cdot\TensorRank{3}{6}{6}}$\\
$\TensorRank{4}{4}{4}$&$\mathrm{49}$&$64$&${\TensorRank{2}{2}{2}\cdot\TensorRank{2}{2}{2}}$\\
$\TensorRank{4}{4}{5}$&$\mathrm{64}$&$80$&${\TensorRank{4}{4}{2}+\TensorRank{4}{4}{3}}$\\
$\TensorRank{4}{4}{6}$&$\mathrm{75}$&$96$&${\TensorRank{4}{4}{4}+\TensorRank{4}{4}{2}}$\\
$\TensorRank{4}{4}{7}$&$\mathrm{87}$&$112$&${\TensorRank{4}{4}{4}+\TensorRank{4}{4}{3}}$\\
$\TensorRank{4}{4}{8}$&$\mathrm{98}$&$128$&${\TensorRank{1}{1}{2}\cdot\TensorRank{4}{4}{4}}$\\
$\TensorRank{4}{4}{9}$&$\mathrm{113}$&$144$&${\TensorRank{4}{4}{2}+\TensorRank{4}{4}{7}}$\\
$\TensorRank{4}{4}{10}$&$\mathrm{124}$&$160$&${\TensorRank{4}{4}{2}+\TensorRank{4}{4}{8}}$\\
$\TensorRank{4}{4}{11}$&$\mathrm{136}$&$176$&${\TensorRank{4}{4}{3}+\TensorRank{4}{4}{8}}$\\
$\TensorRank{4}{4}{12}$&$\mathrm{147}$&$192$&${\TensorRank{1}{1}{3}\cdot\TensorRank{4}{4}{4}}$\\
$\TensorRank{4}{5}{5}$&$\mathrm{80}$&$100$&${\TensorRank{2}{5}{5}\cdot\TensorRank{2}{1}{1}}$\\
$\TensorRank{4}{5}{6}$&$\mathbf{93}$&$120$&${3\,\TensorRank{2}{2}{3}+4\,\TensorRank{2}{3}{3}}$\\
$\TensorRank{4}{5}{7}$&$\mathbf{109}$&$140$&${\TensorRank{2}{2}{3}+2\,\TensorRank{2}{2}{4}+2\,\TensorRank{2}{3}{3}+2\,\TensorRank{2}{3}{4}}$\\
$\TensorRank{4}{5}{8}$&$\mathbf{122}$&$160$&${3\,\TensorRank{2}{2}{4}+4\,\TensorRank{2}{3}{4}}$\\
$\TensorRank{4}{5}{9}$&$\mathbf{140}$&$180$&${\TensorRank{2}{2}{4}+2\,\TensorRank{2}{2}{5}+2\,\TensorRank{2}{3}{4}+2\,\TensorRank{2}{3}{5}}$\\
$\TensorRank{4}{5}{10}$&$\mathbf{154}$&$200$&${3\,\TensorRank{2}{2}{5}+4\,\TensorRank{2}{3}{5}}$\\
$\TensorRank{4}{5}{11}$&$\mathbf{170}$&$220$&${\TensorRank{2}{2}{5}+2\,\TensorRank{2}{2}{6}+2\,\TensorRank{2}{3}{5}+2\,\TensorRank{2}{3}{6}}$\\
$\TensorRank{4}{5}{12}$&$\mathbf{183}$&$240$&${3\,\TensorRank{2}{2}{6}+4\,\TensorRank{2}{3}{6}}$\\
$\TensorRank{4}{6}{6}$&$\mathrm{105}$&$144$&${\TensorRank{2}{2}{2}\cdot\TensorRank{2}{3}{3}}$\\
$\TensorRank{4}{6}{7}$&$\mathbf{125}$&$168$&${3\,\TensorRank{2}{3}{3}+4\,\TensorRank{2}{3}{4}}$\\
$\TensorRank{4}{6}{8}$&$\mathrm{140}$&$192$&${\TensorRank{2}{2}{2}\cdot\TensorRank{2}{3}{4}}$\\
$\TensorRank{4}{6}{9}$&$\mathbf{160}$&$216$&${3\,\TensorRank{2}{3}{4}+4\,\TensorRank{2}{3}{5}}$\\
$\TensorRank{4}{6}{10}$&$\mathrm{175}$&$240$&${\TensorRank{2}{2}{2}\cdot\TensorRank{2}{3}{5}}$\\
$\TensorRank{4}{6}{11}$&$\mathrm{198}$&$264$&${\TensorRank{4}{6}{3}+\TensorRank{4}{6}{8}}$\\
$\TensorRank{4}{6}{12}$&$\mathrm{210}$&$288$&${\TensorRank{1}{1}{2}\cdot\TensorRank{4}{6}{6}}$\\
$\TensorRank{4}{7}{7}$&$\mathbf{147}$&$196$&${\TensorRank{2}{3}{3}+2\,\TensorRank{2}{4}{4}+4\,\TensorRank{2}{3}{4}}$\\
$\TensorRank{4}{7}{8}$&$\mathbf{164}$&$224$&${3\,\TensorRank{2}{3}{4}+4\,\TensorRank{2}{4}{4}}$\\
$\TensorRank{4}{7}{9}$&$\mathbf{188}$&$252$&${\TensorRank{2}{3}{4}+2\,\TensorRank{2}{3}{5}+2\,\TensorRank{2}{4}{4}+2\,\TensorRank{2}{4}{5}}$\\
$\TensorRank{4}{7}{10}$&$\mathbf{207}$&$280$&${3\,\TensorRank{2}{3}{5}+4\,\TensorRank{2}{4}{5}}$\\
$\TensorRank{4}{7}{11}$&$\mathbf{229}$&$308$&${\TensorRank{2}{3}{5}+2\,\TensorRank{2}{3}{6}+2\,\TensorRank{2}{4}{5}+2\,\TensorRank{2}{4}{6}}$\\
$\TensorRank{4}{7}{12}$&$\mathbf{246}$&$336$&${3\,\TensorRank{2}{3}{6}+4\,\TensorRank{2}{4}{6}}$\\
$\TensorRank{4}{8}{8}$&$\mathrm{182}$&$256$&${\TensorRank{2}{2}{2}\cdot\TensorRank{2}{4}{4}}$\\
$\TensorRank{4}{8}{9}$&$\mathbf{210}$&$288$&${3\,\TensorRank{2}{4}{4}+4\,\TensorRank{2}{4}{5}}$\\
$\TensorRank{4}{8}{10}$&$\mathrm{231}$&$320$&${\TensorRank{2}{2}{2}\cdot\TensorRank{2}{4}{5}}$\\
$\TensorRank{4}{8}{11}$&$\mathbf{255}$&$352$&${3\,\TensorRank{2}{4}{5}+4\,\TensorRank{2}{4}{6}}$\\
$\TensorRank{4}{8}{12}$&$\mathrm{273}$&$384$&${\TensorRank{2}{2}{2}\cdot\TensorRank{2}{4}{6}}$\\
$\TensorRank{4}{9}{9}$&$\mathrm{225}$&$324$&${\TensorRank{2}{3}{3}\cdot\TensorRank{2}{3}{3}}$\\
$\TensorRank{4}{9}{10}$&$\mathbf{259}$&$360$&${3\,\TensorRank{2}{4}{5}+4\,\TensorRank{2}{5}{5}}$\\
$\TensorRank{4}{9}{11}$&$\mathrm{284}$&$396$&${\TensorRank{4}{9}{2}+\TensorRank{4}{9}{9}}$\\
$\TensorRank{4}{9}{12}$&$\mathrm{300}$&$432$&${\TensorRank{2}{3}{3}\cdot\TensorRank{2}{3}{4}}$\\
$\TensorRank{4}{10}{10}$&$\mathrm{280}$&$400$&${\TensorRank{2}{2}{2}\cdot\TensorRank{2}{5}{5}}$\\
$\TensorRank{4}{10}{11}$&$\mathrm{320}$&$440$&${\TensorRank{4}{10}{1}+\TensorRank{4}{10}{10}}$\\
$\TensorRank{4}{10}{12}$&$\mathrm{336}$&$480$&${\TensorRank{2}{2}{2}\cdot\TensorRank{2}{5}{6}}$\\
$\TensorRank{4}{11}{11}$&$\mathbf{346}$&$484$&${\TensorRank{2}{5}{5}+2\,\TensorRank{2}{6}{6}+4\,\TensorRank{2}{5}{6}}$\\
$\TensorRank{4}{11}{12}$&$\mathbf{372}$&$528$&${3\,\TensorRank{2}{5}{6}+4\,\TensorRank{2}{6}{6}}$\\
$\TensorRank{4}{12}{12}$&$\mathrm{390}$&$576$&${\TensorRank{2}{3}{3}\cdot\TensorRank{2}{4}{4}}$\\
$\TensorRank{5}{5}{5}$&$99$&$125$&Sedoglavic~\cite{Sedoglavic:2017ab}\\
$\TensorRank{5}{5}{6}$&$\mathbf{117}$&$150$&${\TensorRank{2}{2}{3}+2\,\TensorRank{3}{3}{3}+4\,\TensorRank{2}{3}{3}}$\\
$\TensorRank{5}{5}{7}$&$\mathbf{136}$&$175$&${\TensorRank{2}{2}{4}+\TensorRank{3}{3}{3}+\TensorRank{3}{3}{4}+2\,\TensorRank{2}{3}{3}+2\,\TensorRank{2}{3}{4}}$\\
$\TensorRank{5}{5}{8}$&$\mathbf{152}$&$200$&${\TensorRank{2}{2}{4}+2\,\TensorRank{3}{3}{4}+4\,\TensorRank{2}{3}{4}}$\\
$\TensorRank{5}{5}{9}$&$\mathbf{173}$&$225$&${\TensorRank{2}{2}{5}+\TensorRank{3}{3}{4}+\TensorRank{3}{3}{5}+2\,\TensorRank{2}{3}{4}+2\,\TensorRank{2}{3}{5}}$\\
$\TensorRank{5}{5}{10}$&$\mathbf{190}$&$250$&${\TensorRank{2}{2}{5}+2\,\TensorRank{3}{3}{5}+4\,\TensorRank{2}{3}{5}}$\\
$\TensorRank{5}{5}{11}$&$\mathbf{206}$&$275$&${\TensorRank{2}{2}{6}+\TensorRank{2}{3}{6}+2\,\TensorRank{3}{3}{6}+3\,\TensorRank{2}{3}{5}}$\\
$\TensorRank{5}{5}{12}$&$\mathbf{221}$&$300$&${\TensorRank{2}{2}{6}+2\,\TensorRank{3}{3}{6}+4\,\TensorRank{2}{3}{6}}$\\
$\TensorRank{5}{6}{6}$&$\mathbf{137}$&$180$&${3\,\TensorRank{2}{3}{3}+4\,\TensorRank{3}{3}{3}}$\\
$\TensorRank{5}{6}{7}$&$\mathbf{159}$&$210$&${\TensorRank{2}{3}{3}+2\,\TensorRank{2}{3}{4}+2\,\TensorRank{3}{3}{3}+2\,\TensorRank{3}{3}{4}}$\\
$\TensorRank{5}{6}{8}$&$\mathbf{176}$&$240$&${3\,\TensorRank{2}{3}{4}+4\,\TensorRank{3}{3}{4}}$\\
$\TensorRank{5}{6}{9}$&$\mathbf{200}$&$270$&${\TensorRank{2}{3}{4}+2\,\TensorRank{2}{3}{5}+2\,\TensorRank{3}{3}{4}+2\,\TensorRank{3}{3}{5}}$\\
$\TensorRank{5}{6}{10}$&$\mathbf{218}$&$300$&${\TensorRank{2}{3}{4}+2\,\TensorRank{2}{3}{6}+2\,\TensorRank{3}{3}{4}+2\,\TensorRank{3}{3}{6}}$\\
$\TensorRank{5}{6}{11}$&$\mathbf{236}$&$330$&${\TensorRank{2}{3}{6}+\TensorRank{3}{3}{5}+2\,\TensorRank{2}{3}{5}+3\,\TensorRank{3}{3}{6}}$\\
$\TensorRank{5}{6}{12}$&$\mathbf{250}$&$360$&${3\,\TensorRank{2}{3}{6}+4\,\TensorRank{3}{3}{6}}$\\
$\TensorRank{5}{7}{7}$&$\mathbf{185}$&$245$&${\TensorRank{2}{4}{4}+\TensorRank{3}{3}{3}+\TensorRank{3}{4}{4}+2\,\TensorRank{2}{3}{4}+2\,\TensorRank{3}{3}{4}}$\\
$\TensorRank{5}{7}{8}$&$\mathbf{206}$&$280$&${\TensorRank{2}{3}{4}+2\,\TensorRank{2}{4}{4}+2\,\TensorRank{3}{3}{4}+2\,\TensorRank{3}{4}{4}}$\\
$\TensorRank{5}{7}{9}$&$\mathbf{235}$&$315$&${\TensorRank{2}{3}{5}+\TensorRank{2}{4}{4}+\TensorRank{2}{4}{5}+\TensorRank{3}{3}{4}+\TensorRank{3}{3}{5}+\TensorRank{3}{4}{4}+\TensorRank{3}{4}{5}}$\\
$\TensorRank{5}{7}{10}$&$\mathbf{259}$&$350$&${\TensorRank{2}{3}{5}+2\,\TensorRank{2}{4}{5}+2\,\TensorRank{3}{3}{5}+2\,\TensorRank{3}{4}{5}}$\\
$\TensorRank{5}{7}{11}$&$\mathbf{283}$&$385$&${\TensorRank{2}{3}{5}+\TensorRank{2}{4}{5}+\TensorRank{2}{4}{6}+\TensorRank{3}{4}{5}+\TensorRank{3}{4}{6}+2\,\TensorRank{3}{3}{6}}$\\
$\TensorRank{5}{7}{12}$&$\mathbf{304}$&$420$&${\TensorRank{2}{3}{6}+2\,\TensorRank{2}{4}{6}+2\,\TensorRank{3}{3}{6}+2\,\TensorRank{3}{4}{6}}$\\
$\TensorRank{5}{8}{8}$&$\mathbf{230}$&$320$&${3\,\TensorRank{2}{4}{4}+4\,\TensorRank{3}{4}{4}}$\\
$\TensorRank{5}{8}{9}$&$\mathbf{264}$&$360$&${\TensorRank{2}{4}{4}+2\,\TensorRank{2}{4}{5}+2\,\TensorRank{3}{4}{4}+2\,\TensorRank{3}{4}{5}}$\\
$\TensorRank{5}{8}{10}$&$\mathbf{291}$&$400$&${3\,\TensorRank{2}{4}{5}+4\,\TensorRank{3}{4}{5}}$\\
$\TensorRank{5}{8}{11}$&$\mathbf{323}$&$440$&${\TensorRank{2}{4}{5}+2\,\TensorRank{2}{4}{6}+2\,\TensorRank{3}{4}{5}+2\,\TensorRank{3}{4}{6}}$\\
$\TensorRank{5}{8}{12}$&$\mathrm{346}$&$480$&${\TensorRank{5}{2}{12}+\TensorRank{5}{6}{12}}$\\
$\TensorRank{5}{9}{9}$&$\mathbf{300}$&$405$&${\TensorRank{2}{6}{6}+\TensorRank{3}{3}{3}+\TensorRank{3}{6}{6}+2\,\TensorRank{2}{3}{6}+2\,\TensorRank{3}{3}{6}}$\\
$\TensorRank{5}{9}{10}$&$\mathbf{331}$&$450$&${\TensorRank{2}{4}{5}+2\,\TensorRank{2}{5}{5}+2\,\TensorRank{3}{4}{5}+2\,\TensorRank{3}{5}{5}}$\\
$\TensorRank{5}{9}{11}$&$\mathbf{360}$&$495$&${\TensorRank{2}{3}{5}+\TensorRank{2}{5}{6}+\TensorRank{2}{6}{6}+\TensorRank{3}{5}{6}+\TensorRank{3}{6}{6}+2\,\TensorRank{3}{3}{6}}$\\
$\TensorRank{5}{9}{12}$&$\mathbf{384}$&$540$&${\TensorRank{2}{3}{6}+2\,\TensorRank{2}{6}{6}+2\,\TensorRank{3}{3}{6}+2\,\TensorRank{3}{6}{6}}$\\
$\TensorRank{5}{10}{10}$&$\mathbf{364}$&$500$&${3\,\TensorRank{2}{5}{5}+4\,\TensorRank{3}{5}{5}}$\\
$\TensorRank{5}{10}{11}$&$\mathbf{398}$&$550$&${\TensorRank{2}{5}{6}+\TensorRank{3}{5}{5}+2\,\TensorRank{2}{5}{5}+3\,\TensorRank{3}{5}{6}}$\\
$\TensorRank{5}{10}{12}$&$\mathbf{424}$&$600$&${3\,\TensorRank{2}{5}{6}+4\,\TensorRank{3}{5}{6}}$\\
$\TensorRank{5}{11}{11}$&$\mathbf{434}$&$605$&${\TensorRank{2}{6}{6}+\TensorRank{3}{5}{5}+\TensorRank{3}{6}{6}+2\,\TensorRank{2}{5}{6}+2\,\TensorRank{3}{5}{6}}$\\
$\TensorRank{5}{11}{12}$&$\mathbf{462}$&$660$&${\TensorRank{2}{5}{6}+2\,\TensorRank{2}{6}{6}+2\,\TensorRank{3}{5}{6}+2\,\TensorRank{3}{6}{6}}$\\
$\TensorRank{5}{12}{12}$&$\mathbf{491}$&$720$&${3\,\TensorRank{2}{6}{6}+4\,\TensorRank{3}{6}{6}}$\\
$\TensorRank{6}{6}{6}$&$\mathrm{160}$&$216$&${\TensorRank{3}{3}{6}\cdot\TensorRank{2}{2}{1}}$\\
$\TensorRank{6}{6}{7}$&$\mathbf{185}$&$252$&${3\,\TensorRank{3}{3}{3}+4\,\TensorRank{3}{3}{4}}$\\
$\TensorRank{6}{6}{8}$&$\mathrm{203}$&$288$&${\TensorRank{2}{2}{2}\cdot\TensorRank{3}{3}{4}}$\\
$\TensorRank{6}{6}{9}$&$\mathrm{225}$&$324$&${\TensorRank{2}{3}{3}\cdot\TensorRank{3}{2}{3}}$\\
$\TensorRank{6}{6}{10}$&$\mathbf{247}$&$360$&${3\,\TensorRank{3}{3}{4}+4\,\TensorRank{3}{3}{6}}$\\
$\TensorRank{6}{6}{11}$&$\mathbf{268}$&$396$&${3\,\TensorRank{3}{3}{5}+4\,\TensorRank{3}{3}{6}}$\\
$\TensorRank{6}{6}{12}$&$\mathrm{280}$&$432$&${\TensorRank{2}{2}{2}\cdot\TensorRank{3}{3}{6}}$\\
$\TensorRank{6}{7}{7}$&$\mathbf{215}$&$294$&${\TensorRank{3}{3}{3}+2\,\TensorRank{3}{4}{4}+4\,\TensorRank{3}{3}{4}}$\\
$\TensorRank{6}{7}{8}$&$\mathbf{239}$&$336$&${3\,\TensorRank{3}{3}{4}+4\,\TensorRank{3}{4}{4}}$\\
$\TensorRank{6}{7}{9}$&$\mathbf{273}$&$378$&${\TensorRank{3}{3}{4}+2\,\TensorRank{3}{3}{5}+2\,\TensorRank{3}{4}{4}+2\,\TensorRank{3}{4}{5}}$\\
$\TensorRank{6}{7}{10}$&$\mathbf{300}$&$420$&${3\,\TensorRank{3}{3}{5}+4\,\TensorRank{3}{4}{5}}$\\
$\TensorRank{6}{7}{11}$&$\mathbf{328}$&$462$&${\TensorRank{3}{3}{5}+2\,\TensorRank{3}{3}{6}+2\,\TensorRank{3}{4}{5}+2\,\TensorRank{3}{4}{6}}$\\
$\TensorRank{6}{7}{12}$&$\mathbf{352}$&$504$&${3\,\TensorRank{3}{3}{6}+4\,\TensorRank{3}{4}{6}}$\\
$\TensorRank{6}{8}{8}$&$\mathrm{266}$&$384$&${\TensorRank{2}{2}{2}\cdot\TensorRank{3}{4}{4}}$\\
$\TensorRank{6}{8}{9}$&$\mathrm{300}$&$432$&${\TensorRank{2}{4}{3}\cdot\TensorRank{3}{2}{3}}$\\
$\TensorRank{6}{8}{10}$&$\mathrm{336}$&$480$&${\TensorRank{2}{2}{2}\cdot\TensorRank{3}{4}{5}}$\\
$\TensorRank{6}{8}{11}$&$\mathrm{373}$&$528$&${\TensorRank{6}{2}{11}+\TensorRank{6}{6}{11}}$\\
$\TensorRank{6}{8}{12}$&$\mathrm{390}$&$576$&${\TensorRank{2}{4}{4}\cdot\TensorRank{3}{2}{3}}$\\
$\TensorRank{6}{9}{9}$&$\mathbf{343}$&$486$&${\TensorRank{3}{3}{3}+2\,\TensorRank{3}{6}{6}+4\,\TensorRank{3}{3}{6}}$\\
$\TensorRank{6}{9}{10}$&$\mathrm{375}$&$540$&${\TensorRank{2}{3}{5}\cdot\TensorRank{3}{3}{2}}$\\
$\TensorRank{6}{9}{11}$&$\mathbf{416}$&$594$&${\TensorRank{3}{3}{5}+2\,\TensorRank{3}{3}{6}+2\,\TensorRank{3}{5}{6}+2\,\TensorRank{3}{6}{6}}$\\
$\TensorRank{6}{9}{12}$&$\mathrm{435}$&$648$&${\TensorRank{2}{3}{3}\cdot\TensorRank{3}{3}{4}}$\\
$\TensorRank{6}{10}{10}$&$\mathrm{422}$&$600$&${\TensorRank{6}{10}{4}+\TensorRank{6}{10}{6}}$\\
$\TensorRank{6}{10}{11}$&$\mathbf{463}$&$660$&${3\,\TensorRank{3}{5}{5}+4\,\TensorRank{3}{5}{6}}$\\
$\TensorRank{6}{10}{12}$&$\mathrm{490}$&$720$&${\TensorRank{2}{2}{2}\cdot\TensorRank{3}{5}{6}}$\\
$\TensorRank{6}{11}{11}$&$\mathbf{501}$&$726$&${\TensorRank{3}{5}{5}+2\,\TensorRank{3}{6}{6}+4\,\TensorRank{3}{5}{6}}$\\
$\TensorRank{6}{11}{12}$&$\mathrm{530}$&$792$&${\TensorRank{6}{5}{12}+\TensorRank{6}{6}{12}}$\\
$\TensorRank{6}{12}{12}$&$\mathrm{560}$&$864$&${\TensorRank{2}{2}{2}\cdot\TensorRank{3}{6}{6}}$\\
$\TensorRank{7}{7}{7}$&$\mathbf{250}$&$343$&${\TensorRank{4}{4}{4}+3\,\TensorRank{3}{3}{4}+3\,\TensorRank{3}{4}{4}}$\\
$\TensorRank{7}{7}{8}$&$\mathbf{279}$&$392$&${\TensorRank{3}{3}{4}+2\,\TensorRank{4}{4}{4}+4\,\TensorRank{3}{4}{4}}$\\
$\TensorRank{7}{7}{9}$&$\mathbf{321}$&$441$&${\TensorRank{3}{3}{5}+\TensorRank{4}{4}{4}+\TensorRank{4}{4}{5}+2\,\TensorRank{3}{4}{4}+2\,\TensorRank{3}{4}{5}}$\\
$\TensorRank{7}{7}{10}$&$\mathbf{353}$&$490$&${\TensorRank{4}{4}{4}+2\,\TensorRank{3}{3}{6}+2\,\TensorRank{3}{4}{4}+2\,\TensorRank{4}{4}{6}}$\\
$\TensorRank{7}{7}{11}$&$\mathbf{388}$&$539$&${\TensorRank{4}{4}{5}+2\,\TensorRank{3}{3}{6}+2\,\TensorRank{3}{4}{5}+2\,\TensorRank{4}{4}{6}}$\\
$\TensorRank{7}{7}{12}$&$\mathbf{419}$&$588$&${2\,\TensorRank{3}{3}{6}+2\,\TensorRank{3}{4}{6}+3\,\TensorRank{4}{4}{6}}$\\
$\TensorRank{7}{8}{8}$&$\mathbf{310}$&$448$&${3\,\TensorRank{3}{4}{4}+4\,\TensorRank{4}{4}{4}}$\\
$\TensorRank{7}{8}{9}$&$\mathbf{360}$&$504$&${\TensorRank{3}{4}{4}+2\,\TensorRank{3}{4}{5}+2\,\TensorRank{4}{4}{4}+2\,\TensorRank{4}{4}{5}}$\\
$\TensorRank{7}{8}{10}$&$\mathrm{401}$&$560$&${\TensorRank{7}{8}{2}+\TensorRank{7}{8}{8}}$\\
$\TensorRank{7}{8}{11}$&$\mathrm{441}$&$616$&${\TensorRank{7}{8}{3}+\TensorRank{7}{8}{8}}$\\
$\TensorRank{7}{8}{12}$&$\mathrm{474}$&$672$&${\TensorRank{7}{8}{4}+\TensorRank{7}{8}{8}}$\\
$\TensorRank{7}{9}{9}$&$\mathrm{408}$&$567$&${\TensorRank{3}{9}{9}+\TensorRank{4}{9}{9}}$\\
$\TensorRank{7}{9}{10}$&$\mathbf{454}$&$630$&${\TensorRank{3}{3}{6}+\TensorRank{3}{4}{4}+\TensorRank{3}{6}{6}+\TensorRank{4}{4}{6}+\TensorRank{4}{6}{6}+2\,\TensorRank{3}{4}{6}}$\\
$\TensorRank{7}{9}{11}$&$\mathbf{494}$&$693$&${\TensorRank{3}{3}{6}+\TensorRank{3}{4}{5}+\TensorRank{3}{4}{6}+\TensorRank{3}{5}{6}+\TensorRank{3}{6}{6}+\TensorRank{4}{5}{6}+\TensorRank{4}{6}{6}}$\\
$\TensorRank{7}{9}{12}$&$\mathbf{526}$&$756$&${\TensorRank{3}{3}{6}+2\,\TensorRank{3}{4}{6}+2\,\TensorRank{3}{6}{6}+2\,\TensorRank{4}{6}{6}}$\\
$\TensorRank{7}{10}{10}$&$\mathbf{500}$&$700$&${\TensorRank{3}{6}{6}+\TensorRank{4}{4}{4}+\TensorRank{4}{6}{6}+2\,\TensorRank{3}{4}{6}+2\,\TensorRank{4}{4}{6}}$\\
$\TensorRank{7}{10}{11}$&$\mathbf{545}$&$770$&${\TensorRank{3}{4}{6}+\TensorRank{3}{5}{6}+\TensorRank{3}{6}{6}+\TensorRank{4}{4}{5}+\TensorRank{4}{4}{6}+\TensorRank{4}{5}{6}+\TensorRank{4}{6}{6}}$\\
$\TensorRank{7}{10}{12}$&$\mathbf{578}$&$840$&${\TensorRank{3}{4}{6}+2\,\TensorRank{3}{6}{6}+2\,\TensorRank{4}{4}{6}+2\,\TensorRank{4}{6}{6}}$\\
$\TensorRank{7}{11}{11}$&$\mathbf{590}$&$847$&${\TensorRank{3}{6}{6}+2\,\TensorRank{3}{5}{6}+2\,\TensorRank{4}{5}{5}+2\,\TensorRank{4}{6}{6}}$\\
$\TensorRank{7}{11}{12}$&$\mathbf{626}$&$924$&${\TensorRank{3}{5}{6}+2\,\TensorRank{3}{6}{6}+2\,\TensorRank{4}{5}{6}+2\,\TensorRank{4}{6}{6}}$\\
$\TensorRank{7}{12}{12}$&$\mathbf{660}$&$1008$&${3\,\TensorRank{3}{6}{6}+4\,\TensorRank{4}{6}{6}}$\\
$\TensorRank{8}{8}{8}$&$\mathrm{343}$&$512$&${\TensorRank{2}{2}{2}\cdot\TensorRank{4}{4}{4}}$\\
$\TensorRank{8}{8}{9}$&$\mathrm{400}$&$576$&${\TensorRank{2}{4}{3}\cdot\TensorRank{4}{2}{3}}$\\
$\TensorRank{8}{8}{10}$&$\mathrm{443}$&$640$&${\TensorRank{8}{8}{2}+\TensorRank{8}{8}{8}}$\\
$\TensorRank{8}{8}{11}$&$\mathrm{492}$&$704$&${\TensorRank{8}{8}{4}+\TensorRank{8}{8}{7}}$\\
$\TensorRank{8}{8}{12}$&$\mathrm{520}$&$768$&${\TensorRank{2}{4}{3}\cdot\TensorRank{4}{2}{4}}$\\
$\TensorRank{8}{9}{9}$&$\mathrm{435}$&$648$&${\TensorRank{2}{3}{3}\cdot\TensorRank{4}{3}{3}}$\\
$\TensorRank{8}{9}{10}$&$\mathrm{500}$&$720$&${\TensorRank{2}{3}{5}\cdot\TensorRank{4}{3}{2}}$\\
$\TensorRank{8}{9}{11}$&$\mathrm{551}$&$792$&${\TensorRank{8}{9}{2}+\TensorRank{8}{9}{9}}$\\
$\TensorRank{8}{9}{12}$&$\mathrm{570}$&$864$&${\TensorRank{2}{3}{3}\cdot\TensorRank{4}{3}{4}}$\\
$\TensorRank{8}{10}{10}$&$\mathbf{559}$&$800$&${\TensorRank{4}{4}{4}+2\,\TensorRank{4}{6}{6}+4\,\TensorRank{4}{4}{6}}$\\
$\TensorRank{8}{10}{11}$&$\mathbf{610}$&$880$&${\TensorRank{4}{4}{5}+2\,\TensorRank{4}{4}{6}+2\,\TensorRank{4}{5}{6}+2\,\TensorRank{4}{6}{6}}$\\
$\TensorRank{8}{10}{12}$&$\mathbf{645}$&$960$&${3\,\TensorRank{4}{4}{6}+4\,\TensorRank{4}{6}{6}}$\\
$\TensorRank{8}{11}{11}$&$\mathbf{661}$&$968$&${2\,\TensorRank{4}{5}{5}+2\,\TensorRank{4}{5}{6}+3\,\TensorRank{4}{6}{6}}$\\
$\TensorRank{8}{11}{12}$&$\mathbf{699}$&$1056$&${3\,\TensorRank{4}{5}{6}+4\,\TensorRank{4}{6}{6}}$\\
$\TensorRank{8}{12}{12}$&$\mathrm{735}$&$1152$&${\TensorRank{2}{2}{2}\cdot\TensorRank{4}{6}{6}}$\\
$\TensorRank{9}{9}{9}$&$\mathrm{514}$&$729$&${\textup{Proj}([[0, 0], [2]],\TensorRank{9}{9}{10})}$\\
$\TensorRank{9}{9}{10}$&$\mathrm{540}$&$810$&${\TensorRank{3}{3}{2}\cdot\TensorRank{3}{3}{5}}$\\
$\TensorRank{9}{9}{11}$&$\mathrm{600}$&$891$&${\textup{Proj}([[0, 0], [12]],\TensorRank{9}{9}{12})}$\\
$\TensorRank{9}{9}{12}$&$\mathrm{600}$&$972$&${\TensorRank{3}{3}{2}\cdot\TensorRank{3}{3}{6}}$\\
$\TensorRank{9}{10}{10}$&$\mathrm{625}$&$900$&${\TensorRank{3}{2}{5}\cdot\TensorRank{3}{5}{2}}$\\
$\TensorRank{9}{10}{11}$&$\mathrm{684}$&$990$&${\TensorRank{9}{10}{2}+\TensorRank{9}{10}{9}}$\\
$\TensorRank{9}{10}{12}$&$\mathrm{708}$&$1080$&${\TensorRank{9}{1}{12}+\TensorRank{9}{9}{12}}$\\
$\TensorRank{9}{11}{11}$&$\mathrm{758}$&$1089$&${\TensorRank{9}{11}{2}+\TensorRank{9}{11}{9}}$\\
$\TensorRank{9}{11}{12}$&$\mathrm{768}$&$1188$&${\TensorRank{9}{2}{12}+\TensorRank{9}{9}{12}}$\\
$\TensorRank{9}{12}{12}$&$\mathrm{800}$&$1296$&${\TensorRank{3}{2}{4}\cdot\TensorRank{3}{6}{3}}$\\
$\TensorRank{10}{10}{10}$&$\mathrm{693}$&$1000$&${\TensorRank{2}{2}{2}\cdot\TensorRank{5}{5}{5}}$\\
$\TensorRank{10}{10}{11}$&$\mathbf{765}$&$1100$&${3\,\TensorRank{5}{5}{5}+4\,\TensorRank{5}{5}{6}}$\\
$\TensorRank{10}{10}{12}$&$\mathbf{815}$&$1200$&${\TensorRank{4}{4}{6}+2\,\TensorRank{6}{6}{6}+4\,\TensorRank{4}{6}{6}}$\\
$\TensorRank{10}{11}{11}$&$\mathbf{841}$&$1210$&${\TensorRank{5}{5}{5}+2\,\TensorRank{5}{6}{6}+4\,\TensorRank{5}{5}{6}}$\\
$\TensorRank{10}{11}{12}$&$\mathrm{898}$&$1320$&${\TensorRank{10}{2}{12}+\TensorRank{10}{9}{12}}$\\
$\TensorRank{10}{12}{12}$&$\mathrm{936}$&$1440$&${\TensorRank{2}{4}{4}\cdot\TensorRank{5}{3}{3}}$\\
$\TensorRank{11}{11}{11}$&$\mathbf{922}$&$1331$&${\TensorRank{6}{6}{6}+3\,\TensorRank{5}{5}{6}+3\,\TensorRank{5}{6}{6}}$\\
$\TensorRank{11}{11}{12}$&$\mathrm{977}$&$1452$&${\TensorRank{2}{11}{12}+\TensorRank{9}{11}{12}}$\\
$\TensorRank{11}{12}{12}$&$\mathrm{1022}$&$1584$&${\TensorRank{2}{12}{12}+\TensorRank{9}{12}{12}}$\\
$\TensorRank{12}{12}{12}$&$\mathrm{1040}$&$1728$&${\TensorRank{2}{4}{4}\cdot\TensorRank{6}{3}{3}}$\\
 
$\TensorRank{15}{15}{15}$&$\mathrm{2103}$&$3375$&${\TensorRank{15}{15}{3}+\TensorRank{15}{15}{12}}$ \\
$\TensorRank{18}{18}{18}$&$\mathrm{3200}$&$5832$&${\TensorRank{3}{3}{6}\cdot\TensorRank{6}{6}{3}}$ \\
$\TensorRank{21}{21}{21}$&$\mathrm{5240}$&$9261$&${\TensorRank{12}{12}{12}+3\,\TensorRank{9}{9}{12}+3\,\TensorRank{9}{12}{12}}$ 
\end{longtable}
\end{document}